\documentclass[aps, twocolumn, showpacs, letterpaper]{revtex4}

\usepackage{amsmath}
\usepackage{amssymb}
\usepackage{graphicx}
\usepackage{xspace}
\usepackage[mathscr]{euscript}

\newcommand{\eg}{{e.g.,\/}\xspace}
\newcommand{\ie}{{i.e.,\/}\xspace}

\newcommand{\gap}{\mbox{}}

\newcommand{\eq}[1]{(\ref{#1})}
\newcommand{\Eq}[1]{Eq.~(\ref{#1})}
\newcommand{\Eqs}[1]{Eqs.~(\ref{#1})} 
\newcommand{\Eqsc}[2]{Eqs.~(\ref{#1}), (\ref{#2})}
\newcommand{\Eqsd}[2]{Eqs.~(\ref{#1})-(\ref{#2})}
\newcommand{\Sec}[1]{Sec.~\ref{#1}}

\newcommand{\App}[1]{Appendix~\ref{#1}}

\newcommand{\Ref}[1]{Ref.~\cite{#1}}
\newcommand{\Refs}[1]{Refs.~\cite{#1}}

\newcommand{\citeSec}[2]{[\cite{#1}, #2]}
\newcommand{\RefSec}[2]{Ref.~\citeSec{#1}{#2}}

\newcommand{\mc}[1]{\mathcal{#1}}
\newcommand{\mcc}[1]{\mathfrak{#1}}

\newcommand{\kpt}[1]{{\kern #1 pt}}

\renewcommand{\vec}[1]{{\boldsymbol{\rm #1}}}
\newcommand{\oper}[1]{\hat{\vec{#1}}}
\newcommand{\pd}{\partial}
\newcommand{\const}{\text{const}}
\newcommand{\inv}{\text{inv}}

\newcommand{\transpose}{{\text{T}}}
\newcommand{\act}[1]{\underline{#1}}
\newcommand{\Tr}{\text{Tr}\kpt{1}}
\newcommand{\del}{\nabla}
\newcommand{\sq}{^{\kpt{1}2}}
\newcommand{\osc}[1]{\tilde{#1}}

\newcommand{\db}{{\pd\kpt{.5}\gap'}\kpt{-1}}
\newcommand{\dbt}{\db_t}
\newcommand{\dbdt}{\pd t}

\newcommand{\sN}{N} 												
\newcommand{\sNe}{\sN_e} 										
\newcommand{\N}{\sN} 												
\newcommand{\soscN}{\osc{\sN}}					    

\newcommand{\sV}{V} 												
\newcommand{\vVe}{\vec{\sV}_e} 							
\newcommand{\sv}{V} 												
\newcommand{\vv}{\vec{\sv}} 								
\newcommand{\su}{\osc{V}} 							  	
\newcommand{\vu}{\osc{\vec{V}}} 					  

\newcommand{\sW}{W}                         
\newcommand{\tW}{\oper{\sW}}                
\newcommand{\tWc}{\sW}                      

\newcommand{\sP}{P} 												
\newcommand{\tPe}{\oper{\sP}_e} 						
\newcommand{\tP}{\oper{\sP}} 								
\newcommand{\tPc}{\sP} 											
\newcommand{\tp}{\osc{\tP}} 							  
\newcommand{\tpc}{\osc{\sP}} 							  
\newcommand{\soscp}{\osc{p}}                

\newcommand{\sT}{T}													
\newcommand{\tT}{\oper{\sT}}								
\newcommand{\tTc}{\sT}											
\newcommand{\sC}{C}                         
\newcommand{\tC}{\oper{\sC}}                
\newcommand{\tCc}{\sC}                      

\newcommand{\sE}{E} 										    
\newcommand{\vE}{\vec{\sE}} 						    
\newcommand{\soscE}{\osc{\sE}}              
\newcommand{\voscE}{\osc{\vec{\sE}}}        

\newcommand{\sw}{w}                         
\newcommand{\tw}{\oper{\sw}}                
\newcommand{\swc}{\sw}                      
\newcommand{\vxi}{\vec{\xi}}                

\newcommand{\ssigma}{\sigma}                
\newcommand{\tsigma}{\oper{\sigma}}         
\newcommand{\tsc}{\ssigma}                  

\newcommand{\svph}{v_{\rm ph}}              
\newcommand{\vvg}{\vec{v}_g}                
\newcommand{\vk}{\vec{k}}                   
\newcommand{\sU}{U}                         
\newcommand{\vU}{\vec{\sU}}                 

\newcommand{\sme}{m_e} 											
\newcommand{\smi}{m_i} 											
\newcommand{\sphi}{\varphi} 								
\newcommand{\svte}{v_{Te}}					        
\newcommand{\cmtr}{\{\vu,\vv\}} 						
\newcommand{\permittivity}{\varepsilon_l}   
\newcommand{\nenv}{\mc{N}}                  
\newcommand{\lambdaD}{\lambda_{\rm D}}      

\newcommand{\vrr}{\vec{r}}                  
\newcommand{\vPi}{\vec{\Pi}}                
\newcommand{\sngrad}{h}                     
\newcommand{\vngrad}{\vec{\sngrad}}         

\newcommand{\tI}{\oper{1}}                  
\newcommand{\tH}{\oper{H}}                  
\newcommand{\tG}{\oper{G}}                  

\newcommand{\saa}{A}
\newcommand{\sbb}{B}
\newcommand{\va}{\vec{\saa}}
\newcommand{\vb}{\vec{\sbb}}
\newcommand{\vQ}{\vec{q}}

\begin{document}

\title{Langmuir wave linear evolution in inhomogeneous nonstationary anisotropic plasma}

\author{I.~Y. Dodin, V.~I. Geyko, and N.~J. Fisch}
\affiliation{Department of Astrophysical Sciences, Princeton University, Princeton, New Jersey 08544, USA}
\date{\today}

\begin{abstract}
Equations describing the linear evolution of a non-dissipative Langmuir wave in inhomogeneous nonstationary anisotropic plasma without magnetic field are derived in the geometrical optics approximation. A continuity equation is obtained for the wave action density, and the conditions for the action conservation are formulated. In homogeneous plasma, the wave field $\soscE$ universally scales with the electron density $\N$ as $\soscE \propto \N^{3/4}$, whereas the wavevector evolution varies depending on the wave geometry.
\end{abstract}



\maketitle

\section{Introduction}

The energy of a wave propagating in nonstationary medium can be manipulated by controlling how the parameters of the medium evolve \cite{ref:averkov58, ref:morgenthaler58, ref:stepanov60, ref:ostrovsky71, ref:stepanov93, ref:gildenburg93, ref:bakunov00};  for example, the energy can be pumped up, transported, focused, and (or) deposited where necessary. In the case of Langmuir, or plasma waves \cite{ref:tonks29}, there may be important high energy density applications connected with compressing plasma targets, because these targets may advertently or inadvertently contain wave packets that would be amplified along with the densification. Hence understanding of the Langmuir wave evolution in nonstationary plasma is needed. To develop such understanding is the purpose of this paper.

Specifically, we assume the geometrical optics (GO) limit \cite{book:kravtsov, ref:kravtsov74, ref:bernstein75, ref:bornatici00, ref:bornatici03}, when the plasma parameters vary sufficiently slowly in time and space. We also assume that a wave is linear \cite{foot:nonlin}, and no collisions, ionization, or recombination take place \cite{foot:nonham}. In this case, the plasma dynamics should allow a Lagrangian formulation \cite{ref:low58, foot:lagr}; thus it is anticipated to comply with the general theorem of GO which states that the wave action is conserved in inhomogeneous nonstationary medium \cite{ref:bretherton68, ref:garrett67, ref:whitham65, tex:ostrovsky68, ref:ostrovsky72, ref:katou81, ref:vladimirov96}. Previously, the theorem was independently rederived for a variety of oscillations \cite{ref:bretherton68, ref:garrett67, ref:stepanov67, ref:stepanov68, ref:kravtsov69, ref:stepanov71b}, confirming the general treatment; particularly, space-charge waves in cold electron beams were considered, similar to Langmuir waves in cold plasmas \cite{ref:stepanov63, ref:shevchik59, book:shevchik}. However, for thermal plasmas there has been less agreement, and some of the models proposed in literature do not comply with the action conservation.

The Langmuir wave action, or ``plasmon'' conservation theorem (PCT) was reported in \Refs{ref:vedenov67, ref:bloomberg72, ref:dewar72, ref:zakharov72}, accounting for nonlinear effects; however, the inhomogeneity of the background plasma was neglected there (see also \Ref{ref:mendonca09b}). The density inhomogeneity was included in a linear treatment in \Ref{ref:kravtsov70}, yet within a model assuming constant temperature. More precise models of Langmuir waves in inhomogeneous plasma (see, \eg \Ref{ref:tidman60} and references therein) did not specifically address PCT and assumed stationary medium; also, the wave equation derived in \Ref{ref:parker64} is not entirely correct (see, \eg \Refs{ref:dorman69, ref:peng75} and \Sec{sec:waveeq}) and hence is at variance with the theorem. Similarly, the kinetic models offered in \Refs{ref:bloomberg68, ref:chen71} are erroneous, as explained in \Refs{ref:pikulin79, ref:pikulin73}, and so is the corresponding part of \Ref{ref:bernstein77}, as argued in \App{app:pressure}. Thus, the accuracy of PCT with respect to the temperature corrections was not fully assessed.

An accurate kinetic treatment of the thermal effects was eventually proposed in \Refs{ref:pikulin79, ref:pikulin73}. Particularly, it was shown that the Langmuir wave action in isotropic nonstationary collisionless plasma is conserved in the GO limit, assuming that Landau damping is insignificant. However, the solution in \Refs{ref:pikulin79, ref:pikulin73} is incomplete, because collisionless plasma may not remain isotropic in the presence of inhomogeneous average flow \cite{foot:anisot}. Thus it yet remains to derive an explicit equation for a Langmuir wave in nonstationary inhomogeneous anisotropic plasma and show how the wave parameters evolve.

These results are reported in the present paper, which thus completes the studies in \Refs{ref:stepanov63, ref:kravtsov70, ref:pikulin73, ref:pikulin79} and finally reconciles the Langmuir wave dynamics in inhomogeneous nonstationary warm plasmas with the general principles of the Lagrangian~GO. Specifically, we derive a continuity equation for the wave action density and the explicit conditions under which the action is conserved. Hence it is shown that, in homogeneous plasma carrying a Langmuir wave, the wave field universally scales with the electron density $\N$ as $\soscE \propto \N^{3/4}$, whereas the wavevector evolution varies depending on the wave geometry.

The paper is organized as follows. In \Sec{sec:basic} we introduce our basic equations. In \Sec{sec:homogeneous} we find the Langmuir wave dispersion relation in homogeneous anisotropic plasma. In \Sec{sec:go} we derive the equations for GO rays and the amplitude of Langmuir oscillations in inhomogeneous nonstationary plasma. In \Sec{sec:discuss} we use those to obtain a continuity equation for the wave action density; we also derive the scalings for the oscillation field amplitude and wavenumber. In \Sec{sec:concl} we summarize our main results. Supplementary calculations are given in appendixes.

\section{Basic equations}
\label{sec:basic}

Consider a Langmuir wave in unmagnetized nonrelativistic plasma with given flow velocity $\vv(\vrr, t)$. Neglect ion oscillations and assume collisions and Landau damping to be insignificant on time scales of interest. For electrons adopt the low-temperature approximation requiring $k \lambdaD \ll 1$, where $k$ is the wavenumber, and $\lambdaD$ is the Debye length (cf. Eq.~(11) in \Ref{ref:tokatly99}). Hence an asymptotic closure of the hydrodynamic model is possible, via omitting the heat flux, and one obtains, by taking the first three velocity moments of the electron Vlasov equation \cite{ref:tokatly99, tex:oberman60, ref:bernstein60}:
\begin{gather}
\partial_t \sNe + \del \cdot (\sNe \vVe) = 0, \label{eq:contin}\\
\partial_t \vVe + (\vVe \cdot \del) \vVe = \frac{e}{\sme}\,\vE-\frac{e}{\sme}\,\del\sphi -\frac{\del \cdot \tPe}{\sNe \sme},\label{eq:V}
\end{gather}
\vspace{-.5cm}
\begin{multline}
\partial_t \tPe + (\vVe \cdot \del) \tPe + \tPe (\del \cdot \vVe) + \\
+[(\tPe \del) \vVe] + [(\tPe \del) \vVe]^\transpose = 0.\label{eq:Pe}
\end{multline}
Here $\sNe$ is the electron density, $\vVe$ is the electron flow velocity, $e < 0$ and $\sme$ are the electron charge and mass, $\tPe$ is the electron pressure tensor (which is symmetric by definition), $\vE$ is the average electric field (if any), $\sphi$ is the wave electrostatic potential \cite{foot:nonrel}, and the index $\transpose$ denotes transposition. Separate the slow and the quiver variables, correspondingly, as
\begin{gather}
\sNe = \N + \soscN, \quad \vVe = \vv + \vu, \quad \tPe = \tP + \tp,
\end{gather}
and assume that the oscillations are weak, \ie 
\begin{gather}
n \equiv \soscN/\N \ll 1,
\end{gather}
and similarly for the pressure. (However, the absolute value of $\vv$ is unimportant.) Then the following linear equations are obtained:
\begin{gather}
\dbt n + \vngrad\cdot\vu + \del \cdot \vu = 0, \label{eq:oden}\\
\dbt \vu + (\vu \cdot \del) \vv - \vPi + (e/\sme)\,\del \sphi = 0, \label{eq:odeu}
\end{gather}
\vspace{-.5cm}
\begin{multline}
\dbt \tp + (\vu \cdot \del)\tP + \tp\kpt{2}(\Tr \tW) + (\tW \tp) + (\tW \tp)^\transpose +\\
+ \tP (\del \cdot \vu) + [(\tP \del) \vu] + [(\tP \del) \vu]^\transpose = 0,\label{eq:odep}
\end{multline}
\begin{gather}
\vPi = (n \del \cdot \tP - \del \cdot \tp)/(\sme\N),\label{eq:pidef}\\
\del^2 \sphi = - (\sme/e)\, \omega_p^2 n.\label{eq:odephi}
\end{gather}
Here we introduced the partial time derivative in the frame of reference $K'$ (further denoted by prime) moving with velocity $\vv$ with respect to the laboratory frame $K$:
\begin{gather}
\dbt = \partial_t  + (\vv \cdot \del).
\end{gather}
In addition, we introduced $\vngrad = \del \ln \N$, $\tW = \del \vv$ (which is a tensor with elements $\sW_{j\ell} = \pd \sv_j/\pd x_\ell$), the electron plasma frequency $\omega_p = \sqrt{4\pi e^2 \N/\sme}$, and $\Tr$ for the tensor trace. The complex notation is also henceforth assumed for the quiver variables.

\section{Homogeneous stationary plasma}
\label{sec:homogeneous}

In the case of homogeneous plasma with $\vv = \const$, the exact eigenmodes of \Eqsd{eq:oden}{eq:odephi} are found as follows. Assume
\begin{gather}
n,\kpt{5}\vu,\kpt{5}\tp \sim \exp(i\vk\cdot\vrr - i\omega t).
\end{gather}
Then one gets
\begin{gather}
-i\omega'n + i \vk \cdot \vu = 0, \label{eq:uoden}\\
-i\omega' \vu + i\vk(e/\sme)\,\sphi - \vPi = 0, \label{eq:uodeu}\\
-i\omega'\tp + i\tP (\vk \cdot \vu) + i[(\tP \vk) \vu] + i[(\tP \vk) \vu]^\transpose = 0,\label{eq:uodep}\\
\vPi = - i \vk \tp/(\sme\N),\label{eq:pi2}\\
-k^2 \sphi = - (\sme/e)\, \omega_p^2 n,\label{eq:phi2}
\end{gather}
with $\omega' \equiv \omega - \vk \cdot \vv$ being the wave frequency in the frame where the plasma average flow rests. Hence 
\begin{gather}\label{eq:hv}
\tH \vu = \omega'^2 \vu,
\end{gather}
\ie $\vu$ must be an eigenvector of the tensor $\tH$ given~by
\begin{gather}
\tH = \frac{\vk\vk}{k^2}\,\omega_p^2 + \frac{1}{\sme}\,\Big[2(\tT\vk)\vk + (\vk\cdot \tT\vk)\tI\Big],
\end{gather}
where $\vk\vk$ is a dyad, $\tI$ is a unit tensor, and $\tT = \tP/\N$ is the electron unperturbed temperature tensor.

For a longitudinal wave one has from \Eq{eq:uoden} that
\begin{gather}\label{eq:parv}
\vu = n\omega'\vk/k^2.
\end{gather}
Substituting this into \Eq{eq:hv} yields
\begin{gather}
\tT\vk = \frac{\sme}{2k^2}\Bigg[\omega'^2-\omega_p^2 - \frac{(\vk\cdot \tT\vk)}{\sme}\Bigg]\vk.
\end{gather}
Since the coefficient on the right-hand side is a scalar, $\vk$ must be an eigenvector of~$\tT$:
\begin{gather}
\tT \vk = \sme \svte^2 \vk,
\end{gather}
where the eigenvalues satisfy $\svte^2 > 0$, because $\tT$ is positive defined; thus $\svte$ can then be understood as the electron thermal speed along $\vk$. Then
\begin{gather}\label{eq:uniom2}
\omega = \Omega(\vec{k})  + \vk \cdot \vv,
\end{gather}
where the function $\Omega(\vk)$ (numerically equal to $\omega'$) is given~by $\Omega^2 = \omega_p^2 + 3 k^2 \svte^2$ (cf., \eg \RefSec{book:stix}{Chap.~8}, \Refs{book:landau10, ref:bernstein64, ref:bohm49} for isotropic plasma), or
\begin{gather}\label{eq:Om}
\Omega^2 = \omega_p^2 + \vk \cdot \tC\vk, \quad \tC=3\tT/\sme.
\end{gather}

Therefore, similarly to elastic media \cite{book:iesan, ref:borgnis55,  ref:truesdell61, ref:truesdell66, ref:kolodner66}, a longitudinal wave in plasma must have $\vk$ along a principal axis of the temperature tensor \cite{foot:aniswaves}. Since $\tT$ is symmetric, there always exist at least three such directions. On the other hand, in isotropic plasma, any axis can be considered a principle axis of $\tT$; then Langmuir waves can propagate in arbitrary direction.

\section{Geometrical optics equations}
\label{sec:go}

\subsection{Approximate wave equation}
\label{sec:waveeq}

Consider now a more general case of inhomogeneous nonstationary plasma. Assume, however, that the wave amplitude, the frequency, and local wavevector vary on \textit{large} temporal and spatial scales $\mc{T}$ and $L$ in the plasma average flow rest frame $K'$. Specifically, we assume
\begin{gather}\label{eq:epsilon}
\epsilon \equiv 1/\min \{\omega' \mc{T}, k' L\} \ll 1,
\end{gather}
with $k'=k$ in the nonrelativistic limit used here. In this case, henceforth referred as the geometrical optics (GO) limit \cite{book:kravtsov, ref:kravtsov74}, an approximate scalar equation for a Langmuir wave can be obtained which will capture effects to the leading order in the parameter $\epsilon$.

To derive this equation, first, apply $\dbt$ to \Eq{eq:oden} to~get
\begin{gather}\label{eq:oden2}
\dbt\sq n + \dbt\,(\vngrad\cdot\vu) + \dbt\,(\del \cdot \vu) = 0.
\end{gather}
The third term on the left-hand side here is found by taking the divergence of \Eq{eq:odeu}:
\begin{gather}
\dbt\,(\del \cdot \vu) = \omega_p^2 n + \del \cdot \vPi - (\vu \cdot \del)(\del \cdot \vv) - \cmtr,\nonumber
\end{gather}
where we substituted \Eq{eq:odephi} and introduced
\begin{multline}
\cmtr \equiv \del \cdot \big[
(\vu \cdot \del) \vv + (\vv \cdot \del) \vu \big] - \\
 - (\vu \cdot \del) (\del \cdot \vv) - (\vv \cdot \del)(\del \cdot \vu),
\end{multline}
accounting for the fact that $\dbt$ and $\del$ may not commute. Therefore \Eq{eq:oden2} takes the form
\begin{gather}\label{eq:odenex}
\dbt\sq n + \omega_p^2 n + \del \cdot \vPi + R = 0,
\end{gather}
with $R$ given by
\begin{gather}\label{eq:rdef}
R = - (\vu \cdot \del)(\del \cdot \vv) + \dbt\,(\vngrad\cdot\vu) - \cmtr. 
\end{gather}

The first term in \Eq{eq:rdef} is of order $\epsilon^2$ and can be neglected. The second term is evaluated as
\begin{gather}
\frac{\db (\vngrad\cdot\vu)}{\dbdt}\, \approx 2\,\frac{\db n}{\dbdt}\, \frac{\Omega}{\omega_p}\,\del \omega_p \cdot \frac{\vk}{k^2}
\end{gather}
($\db/\dbdt \equiv \dbt$), where we used that, to the zeroth order in~$\epsilon$, one can employ \Eq{eq:parv} and
\begin{gather}\label{eq:apprderiv}
\dbt \approx - i \Omega, \quad \del \approx i \vk.
\end{gather}
The third term in \Eq{eq:rdef} can be put as (\App{app:comm})
\begin{gather}
\cmtr = 2 (\del \su_j) \cdot (\del \sv_j) - 2 (\del \times \vu) \cdot (\del \times \vv),
\end{gather}
where we assume summation over repeated indexes. Both curls are of order $\epsilon$ here; thus $\cmtr \approx 2 (\del \su_j) \cdot (\del \sv_j)$. \Eqsc{eq:parv}{eq:apprderiv} further yield
\begin{gather}
\cmtr \approx - 2\, (\dbt n)\, \vk\tW\cdot\vk/k^2.
\end{gather}
Hence \Eq{eq:rdef} rewrites as
\begin{gather}\label{eq:r2}
R = 2\,\frac{\db n}{\dbdt} \left( \frac{\Omega}{\omega_p}\,\del \omega_p  + \vk\tW \right) \cdot \frac{\vk}{k^2}.
\end{gather}

An approximate expression for $\del \cdot \vPi$ is obtained similarly (\App{app:pressure}). Substituting that and \Eq{eq:r2} into \Eq{eq:odenex} one gets for isotropic plasma \cite{foot:parker}:
\begin{widetext}
\begin{gather}\label{eq:nmainiso}
\frac{\db\sq n}{\dbdt^2} 
+ \omega_p^2 n - 3\svte^2 \del^2 n 
+ 2\,\frac{\db n}{\dbdt} \left( \frac{\Omega}{\omega_p}\,\del \omega_p  + \vk\tW \right) \cdot \frac{\vk}{k^2}
- 6 \del n \cdot \del \svte^2 = 0,
\end{gather}
and in the general case of anisotropic plasma, which we study below,
\begin{gather}\label{eq:nmain}
\frac{\db\sq n}{\dbdt^2} 
+ \omega_p^2 n 
- \tCc_{j \ell}\,\frac{\pd^2 n}{\pd x_j \, \pd x_\ell}
+ 2\,\frac{\db n}{\dbdt} \left( \frac{\Omega}{\omega_p}\,\frac{\pd \omega_p}{\pd x_\ell}  + k_j\tWc_{j \ell} \right) \frac{k_\ell}{k^2} 
-\left(\delta_{js} + \frac{k_jk_s}{k^2}\right)\frac{\pd \tCc_{s\ell}}{\pd x_j}\,\frac{\pd n}{\pd x_\ell} = 0.
\end{gather}
\end{widetext}

\subsection{Eikonal equation}
\label{sec:rays}

Equation~\eq{eq:nmain} can be solved using the GO approach \cite{ref:stepanov96, book:kravtsov}, specifically as follows. Take
\begin{gather}\label{eq:env}
n = \nenv e^{i \theta},
\end{gather}
where $\nenv$ is the slowly varying envelope. Substitute \Eq{eq:env} into \Eq{eq:nmain} and first consider the terms of order $\epsilon^0$; hence the eikonal equation
\begin{gather}\label{eq:eik}
[-(\pd_t \theta - \del \theta \cdot\vv)^2 + \omega^2_p + \del \theta \cdot \tC \del \theta]\,\nenv= 0.
\end{gather}
Since, by definition,
\begin{gather}\label{eq:omk}
\pd_t \theta = -\omega, \quad \del \theta = \vk,
\end{gather}
\Eq{eq:eik} is equivalent to \Eq{eq:uniom2}, except now the plasma parameters may slowly depend on $\vrr$ and $t$:
\begin{gather}\label{eq:uniom3}
\omega = \Omega(\vk; \vrr, t)  + \vk \cdot \vv(\vrr, t).
\end{gather}
Differentiate \Eq{eq:uniom3} with respect to $t$ and with respect to $\vrr$ and use $\del \omega = - \pd_t \vk$, flowing from \Eqs{eq:omk}. Then
\begin{gather}
d_t \omega = \pd_t \omega(\vk, \vrr, t), \quad
\quad d_t \vk = - \del \omega(\vk, \vrr, t),\label{eq:rays}
\end{gather}
where the partial derivatives are taken at fixed $\vk$;~also
\begin{gather}\label{eq:dt}
d_t \equiv \pd_t + (\vvg \cdot \del),
\end{gather}
and $\vvg \equiv \pd_\vk \omega(\vk, \vrr, t)$ is the group velocity:
\begin{gather}\label{eq:vg2}
\vvg = \vU + \vv, \quad \vU \equiv \pd_\vk \Omega(\vk, \vrr, t).
\end{gather}
Since $\vvg$ equals the velocity at which the envelope propagates \citeSec{book:stix}{Chap.~4}, one can also write
\begin{gather}\label{eq:drdt}
d_t \vrr = \pd_\vk\omega.
\end{gather}
Together, \Eqsc{eq:rays}{eq:drdt} are known as GO ray equations \citeSec{book:stix}{Chap.~4} and can be considered as canonical equations [with the Hamiltonian $\hbar\omega(\vk, \vrr, t)$; \Eq{eq:uniom3}] which determine the dynamics of ``plasmons'', \ie quasiparticles with velocity $\vvg$, momentum $\hbar\vk$, and energy~$\hbar\omega$; see also \Ref{ref:vedenov67, ref:bloomberg72}.

\subsection{Amplitude equation}
\label{sec:ampleq}

The equation obtained from \Eq{eq:nmain} in the first order in $\epsilon$ reads
\begin{widetext}
\begin{gather}\label{eq:ampl1}
-2i\Omega\, \frac{\db \nenv}{\dbdt}-i\frac{\db \Omega}{\dbdt}\,\nenv
-iC_{j\ell}\left(k_\ell\, \frac{\pd \nenv}{\pd x_j}+ k_j\,\frac{\pd \nenv}{\pd x_\ell}\right)
-iC_{j\ell}\nenv\,\frac{\pd k_\ell}{\pd x_j}
-2i\Omega\,\Gamma\nenv
-ik_\ell\nenv\,\frac{\pd C_{s\ell}}{\pd x_j}\left(\delta_{js} + \frac{k_jk_s}{k^2}\right)
 = 0,
\end{gather}
where $\vk$ is a function of $\vrr$ and $t$ [unlike in \Eqs{eq:rays}, where $\vk$ is an independent variable], and
\begin{gather}\label{eq:Gamma}
\Gamma = \left( \frac{\Omega}{\omega_p}\,\frac{\pd \omega_p}{\pd x_\ell}  + k_j\tWc_{j \ell} \right) \frac{k_\ell}{k^2}.
\end{gather}
The same expression can be written also as follows. Use \Eq{eq:rays} for $d_t \vk$ to get
\begin{gather}\label{eq:veck}
\frac{d k_\ell}{dt} = - \frac{\omega_p}{\Omega}\,\frac{\pd \omega_p}{\pd x_\ell} - \frac{k_j k_s}{2\Omega}\,\frac{\partial C_{j s}}{\pd x_\ell} -k_j\tWc_{j \ell}.
\end{gather}
Hence \Eq{eq:Gamma} is put in the form
\begin{gather}\label{eq:Gammab}
\Gamma = - \frac{k_\ell}{k^2}\,\frac{dk_\ell}{dt} +
\left(\frac{\Omega}{\omega_p}-\frac{\omega_p}{\Omega}\right)\frac{k_\ell}{k^2}\,\frac{\pd\omega_p}{\pd x_\ell}
- \frac{k_j k_\ell k_s}{2\Omega k^2}\,\frac{\pd C_{js}}{\pd x_\ell}.
\end{gather}
We now use the expression for $\Omega$ [\Eq{eq:Om}], $\del \omega_p/ \omega_p = \del \N/(2\N)$, and
\begin{gather}\label{eq:scalk}
 \vk \cdot d_t\vk/k = d_t k.
\end{gather}
Hence \Eq{eq:Gamma} can be represented as
\begin{gather}\label{eq:Gamma2}
\Gamma = - \frac{d\ln k}{dt} 
+ \frac{k_jk_\ell k_s}{2\Omega k^2}\left(
C_{js}\sngrad_\ell
- \frac{\pd C_{js}}{\pd x_\ell}
\right).
\end{gather}
Since the wave is assumed propagating along a local principal axis of the temperature tensor \cite{foot:notlang}, one also has $C_{j\ell} k_j k_\ell/k^2  = 3\svte^2$ and $3\svte^2 k_s = C_{s\ell} k_\ell$; thus \Eq{eq:Gamma2} can be further put as
\begin{gather}\label{eq:Gamma3}
\Gamma = - \frac{d\ln k}{dt} 
+ \frac{\sU_j\sngrad_j}{2}
- \frac{k_jk_\ell k_s}{2\Omega k^2}\frac{\pd C_{j\ell}}{\pd x_s},
\end{gather}
where we used $\sU_j = C_{j \ell} k_\ell/\Omega$ [\Eq{eq:vg2}]. Then \Eq{eq:ampl1} rewrites~as
\begin{gather}\label{eq:action0}
\dbt(\Omega |\nenv|^2) + \del \cdot (\Omega|\nenv|^2\vU) + \Omega|\nenv|^2\,(\vU\cdot\vngrad - d_t \ln k^2)= 0.
\end{gather}

We now use
\begin{gather}
\dbt(\Omega |\nenv|^2) + \del \cdot (\Omega|\nenv|^2\vU) = d_t (\Omega |\nenv|^2) + \Omega|\nenv|^2 \del \cdot \vU, \quad
\vU \cdot \vngrad = d_t \ln\N + \del \cdot \vv,
\end{gather}
\end{widetext}
the latter being due to
\begin{gather}\label{eq:slown}
\dbt \N = - \N \del \cdot \vv,
\end{gather}
which flows from \Eq{eq:contin}. Hence \Eq{eq:action0} reads
\begin{gather}\label{eq:action1}
d_t \ln(\Omega |\nenv|^2) + d_t \ln\N - d_t \ln k^2 + \del \cdot \vvg = 0.
\end{gather}
In principle, this allows one to calculate the envelope amplitude $|\nenv|$ along the GO rays, as discussed in \Sec{sec:action}.

\section{Discussion}
\label{sec:discuss}

\subsection{Wave action. Equation of state}
\label{sec:action}

Introduce the wave average energy density
\begin{gather}\label{eq:end}
\mc{E}' = \frac{|\soscE'|^2}{16\pi}\,\frac{\pd (\permittivity' \omega')}{\pd \omega'}
\end{gather}
in the frame $K'$ traveling with velocity $\vv$. Here $\voscE' \approx \voscE$, and the longitudinal permittivity in $K'$ equals that in the laboratory frame $K$: $\permittivity' = \permittivity$ \cite{ref:lichtenberg64}. Neglecting the corrections due to finite $\epsilon$ and using \Eqsd{eq:oden}{eq:pi2}, $\permittivity$ is derived like for isotropic plasma \citeSec{book:stix}{Chap.~3}:
\begin{gather}
\permittivity = 1 -\frac{\omega_p^2}{\omega'^2-3k^2\svte^2}.
\end{gather}
(Here $\vk$ parallel to the principal axis of $\tT$ is assumed, as before \cite{foot:notlang}.) Thus \Eq{eq:end} rewrites~as
\begin{gather}
\mc{E}' = \frac{\Omega^2}{\omega_p^2}\,\frac{|\soscE|^2}{8\pi}
\end{gather}
[where we used ${\permittivity(\Omega+ \vk \cdot \vv; \vk)=0}$], so \textit{this} energy density is always positive, unlike that in $K$~\cite{ref:nezlin76}. 

Further, define the wave action density $J$, or the number of quanta (plasmons) per unit volume, as $J = \mc{E}'/\omega'$ \cite{ref:bretherton68, ref:dewar77}, where we take $\omega'>0$, by analogy with discrete systems (see, \eg Sec.~III of \Ref{my:nlinphi}). Then
\begin{gather}\label{eq:j2}
J \propto \Omega|\nenv|^2\N/k^2,
\end{gather}
where we used $\voscE \approx - i\vk  \sphi$ and \Eq{eq:phi2} for $\sphi$. From \Eq{eq:action1}, it follows then that $d_t \ln J + \del \cdot \vvg = 0$, or
\begin{gather}\label{eq:action2}
d_t J + J \del \cdot \vvg = 0.
\end{gather}
The latter is also equivalent to a continuity equation:
\begin{gather}\label{eq:action3}
\pd_t J + \del \cdot (\vvg J) = 0.
\end{gather}
Hence the wave total action is conserved,
\begin{gather}\label{eq:actionint}
\int J\,d^3\vrr = \inv,
\end{gather}
and the dynamics is thereby called adiabatic.

\Eqsd{eq:action2}{eq:actionint} agree with the previous results for space-charge waves in cold plasmas and electron beams \cite{ref:stepanov63, ref:shevchik59, book:shevchik}, as well as phenomenological hydrodynamical treatment of the corrections due to the electron homogeneous temperature \cite{ref:kravtsov70} and kinetic treatment for inhomogeneous nonstationary but isotropic plasmas \cite{ref:pikulin79, ref:pikulin73}. By construction \cite{ref:tokatly99, tex:oberman60, ref:bernstein60}, the hydrodynamic calculation offered here is asymptotically precise at small temperatures and, apart from missing Landau damping (\Sec{sec:cond}), just as accurate as the perturbative kinetic calculation in \Refs{ref:pikulin79, ref:pikulin73}. On the other hand, it also accounts for the temperature anisotropy, which is anticipated at collisionless compression or rarefaction \cite{foot:anisot} yet missed in \Refs{ref:pikulin79, ref:pikulin73}. Therefore, our results complete those in \Refs{ref:pikulin79, ref:pikulin73} and finally reconcile the Langmuir wave dynamics (particularly the temperature effects; cf. \eg \Refs{ref:bloomberg68, ref:chen71, ref:parker64}) with the general action conservation theorem, which is supposed to hold for any Lagrangian waves \cite{ref:bretherton68, ref:garrett67, ref:whitham65, ref:dewar77, ref:kravtsov74, ref:kravtsov69}. 

Besides, the above results show explicitly how the Langmuir wave parameters evolve. Consider, for instance, homogeneous plasma, assuming that the envelope shape remains fixed. Then \Eq{eq:actionint} rewrites as 
\begin{gather}\label{eq:jn}
J/\N = \inv,
\end{gather}
where we used that the total number of electrons is conserved. In the absence of Landau damping (\Sec{sec:cond}) one has $k \svte \ll \omega_p$, and therefore
\begin{gather}\label{eq:coldj}
J \approx |\soscE|^2/(8\pi \omega_p).
\end{gather}
Together with \Eq{eq:jn}, this yields (like in \Ref{ref:stepanov63})
\begin{gather}\label{eq:scale}
\soscE = \soscE_0 (\N/\N_0)^{3/4},
\end{gather}
where the index 0 denotes the initial values. Thus the wave field increases when the plasma is compressed and decreases when the plasma is rarefied.

Finally, \Eq{eq:scale} also results in an effective adiabatic index $\gamma$ for the ponderomotive pressure $p_E$. To see this, consider the known expression for $p_E$ \cite{ref:kentwell87}, using that the field $\voscE$ oscillates at the frequency $\omega = \omega_p$:
\begin{gather}\label{eq:peff}
p_E = |\soscE|^2/(16\pi),
\end{gather}
[In fact, \Eq{eq:peff} itself is also derivable from \Eq{eq:jn}, as shown in \App{app:eos}.] Hence, from \Eq{eq:scale}, one obtains
\begin{gather}\label{eq:peff2}
p_E = \frac{|\soscE_0|^2}{16\pi}\,\left(\frac{\N}{\N_0}\right)^{\!3/2}\!\!.
\end{gather}
Therefore, for the ponderomotive pressure one has ${\gamma = 3/2}$, which is different, say, from ${\gamma = (D+2)/D}$ for the kinetic pressure of $D$-dimensional thermal electron gas without a wave~\cite{book:landau5}.

\subsection{Wavevector. Adiabaticity conditions}
\label{sec:cond}

The scalings \eq{eq:jn}-\eq{eq:peff2} hold for any wave geometry, whereas the dependence of the frequency and the wavevector on plasma parameters may vary, as governed by \Eq{eq:rays}. Particularly, $\omega$ is conserved only in stationary medium, and the dynamics of $\vk$ is discussed below.

\subsubsection{Homogeneous plasma}

To illustrate the evolution of $\vk$, first consider plasma compression such that $\N$ remains homogeneous [which is possible at homogeneous yet not necessarily zero $\del \cdot \vv$; see \Eq{eq:slown}]. Then the wavevector is conserved if $\vv$ is transverse to $\vk$, an example being radial compression of a cylindrical plasma column with $\vk$ along the axis of symmetry. However, if $\vk$ has a component along $\vv$, the wavevector will evolve; specifically,
\begin{gather}\label{eq:kexp}
k =k_0 \exp\Big[{\textstyle \int_0^t} \nu(t')\,dt'\Big],
\end{gather}
with \Eqsc{eq:veck}{eq:scalk} yielding $\nu=\nu_V$,
\begin{gather}
\nu_V = -\vk \cdot \tW\vk/k^2 \sim \sv/L_\sv,
\end{gather}
where $L_\sv$ is the spatial scale on which the compression takes place. For instance, radial compression with $\vv = \chi(t)\kpt{1} \vec{r}$ and $\vk$ along $\vv$ in spherical, cylindrical, and linear geometry equally yield $\nu_V = \chi$.

\subsubsection{Inhomogeneous plasma}

As the next step, consider inhomogeneous plasma, for now assuming $\vv = 0$. In this case $k$ can increase or decrease, depending on $\vk_0$ as well as the density and temperature gradients, so \Eq{eq:kexp} holds with ${\nu=\nu_\del}$,
\begin{gather}\label{eq:nudel}
\nu_\del = -\vk \cdot \del \Omega/k^2 \sim \svph/L_\Omega,
\end{gather}
with $\svph \approx \omega_p/k$ being the phase speed and $L_\Omega \equiv \Omega/|\del \Omega|$ being the characteristic spatial scale.

First, suppose that $k$ increases. Then, on the time scale of order $\tau_e \sim L_\Omega/\svte$, the wavelength becomes comparable to the Debye length $\lambdaD = \svte/\omega_p$, regardless of whether the plasma inhomogeneity is due to the density or the temperature; hence the wave decays because of Landau damping (see also \Refs{ref:fahleen09, ref:bloomberg68}). In other words, dissipation is negligible only at 
\begin{gather}\label{eq:ld}
t \lesssim L_\Omega/\svte.
\end{gather}
Thus, when compression is added, it will proceed adiabatically only if $\nu_\sv \tau_e \gtrsim 1$,~or
\begin{gather}\label{eq:vcond1}
\sv/L_\sv \gtrsim \svte/L_\Omega.
\end{gather}

Assuming that the plasma average flow is entirely controlled by the large ion mass $\smi \gg \sme$ \cite{foot:b}, one can rewrite $\sv$ in \Eq{eq:vcond1} as follows. Express $\vE$ from \Eq{eq:V} and substitute it into a similar equation for ions; hence
\begin{gather}
\dbt \vv \approx - \del \cdot \tP_\Sigma/(\smi \N),
\end{gather}
where we neglected the electron inertia and introduced the total kinetic pressure $\tP_\Sigma \sim \N\sT$ \cite{foot:ion}. Use $\dbt \vv \sim \sv^2/L_\sv$, yielding
\begin{gather}\label{eq:vest2}
\sv \sim c_s \sqrt{L_\sv/L_\Omega},
\end{gather}
where $c_s \sim \svte\sqrt{\sme/\smi}$ is the ion sound speed. Hence \Eq{eq:vcond1} rewrites as
\begin{gather}\label{eq:unif}
L_\Omega \gtrsim (\smi/\sme)\,L_\sv.
\end{gather}
Therefore, only weakly inhomogeneous plasma can be compressed adiabatically when $k$ grows; otherwise a significant percentage of the wave energy is transformed into the particle thermal energy.

Suppose now that $k$ decreases. In this case the envelope approximation holds only on time
\begin{gather}\label{eq:ld2}
t \lesssim \nu^{-1}_\del,
\end{gather}
after which the wavenumber becomes zero, and thus the wave action is no longer conserved. Therefore adiabatic compression must satisfy $\nu_V \gtrsim \nu_\del$, or
\begin{gather}
V/L_\sv \gtrsim \svph/L_\Omega.
\end{gather}
Assuming \Eq{eq:vest2}, this condition hence reads as
\begin{gather}\label{eq:cdk}
L_\Omega \gtrsim \frac{L_\sv}{(k \lambdaD)^2}\,\frac{\smi}{\sme},
\end{gather}
which requires that plasma be even more homogeneous than in the case when $k$ increases [cf. \Eq{eq:unif}].

\section{Conclusions}
\label{sec:concl}

In this paper we show how a non-dissipative Langmuir wave evolves adiabatically in warm unmagnetized inhomogeneous nonstationary plasma. The hydrodynamic calculation offered here is asymptotically precise at small temperatures ($k \lambdaD \ll 1$) and, apart from missing Landau damping, just as accurate as the perturbative kinetic calculation in \Refs{ref:pikulin79, ref:pikulin73}. On the other hand, it also accounts for the temperature anisotropy, which is anticipated at collisionless compression or rarefaction yet missed in \Refs{ref:pikulin79, ref:pikulin73}. Therefore, our results complete those in \Refs{ref:stepanov63, ref:kravtsov70, ref:pikulin73, ref:pikulin79} and finally reconcile the Langmuir wave dynamics in inhomogeneous warm plasmas (cf. \Refs{ref:bloomberg68, ref:chen71, ref:parker64}) with the general principles of the Lagrangian geometrical optics. 

Specifically, we derive a continuity equation [\Eq{eq:action3}] for the wave action density, as well as the explicit conditions under which the wave action is conserved. Hence it is shown that, in homogeneous plasma carrying a Langmuir wave, the wave field universally scales with the electron density as $\soscE \propto \N^{3/4}$. We also show that the wavevector evolution varies depending on the wave geometry. Particularly, during compression $\vk$ is conserved when aligned with the average velocity $\vv(\vrr, t)$ at homogeneous density and temperature, but otherwise changes, with its absolute value following by \Eqsd{eq:kexp}{eq:nudel}. Also, the wave frequency $\omega$ is conserved only when the plasma is stationary, but otherwise evolves according to \Eq{eq:rays}.

\section{Acknowledgments}

This work was supported by the NNSA under the SSAA Program through DOE Research Grant No.~DE-FG52-04NA00139.

\appendix

\begin{widetext}

\section{Auxiliary vector identity}
\label{app:comm}

In this appendix we derive an alternative form of
\begin{gather}\label{eq:ap3}
\{\va, \vb\} \equiv \del \cdot \big[(\va \cdot \del) \vb + (\vb \cdot \del) \va \big] - (\va \cdot \del) (\del \cdot \vb) - (\vb \cdot \del)(\del \cdot \va)
\end{gather}
for two arbitrary fields $\va$ and $\vb$. First, the expression in the square brackets above rewrites as [\cite{book:korn}, Sec.~5.5-2]
\begin{gather}
(\va \cdot \del) \vb + (\vb \cdot \del) \va = \del (\va \cdot \vb) - \va \times (\del \times \vb) - \vb \times (\del \times \va);
\end{gather}
thus its divergence equals
\begin{gather}\label{eq:ap2}
\del \cdot \big[(\va \cdot \del) \vb + (\vb \cdot \del) \va \big] = \del^2 (\va \cdot \vb) - \del \cdot [\va \times (\del \times \vb)] - \del \cdot [\vb \times (\del \times \va)].
\end{gather}
The chain rule for the second term on the right-hand side yields
\begin{gather}
\del \cdot [\va \times (\del \times \vb)] = \del \cdot [\act{\va} \times (\del \times \vb)] + \del \cdot [\va \times (\act{\del \times \vb})],
\end{gather}
where underlined are the vectors to which the differentiation by the external $\del$ applies. Because of the symmetry properties of the scalar triple product [\cite{book:korn}, Sec.~5.2-8] (here of the vectors $\del$, $\va$, and $\del \times \vb$), this also rewrites as
\begin{gather}\label{eq:ap1}
\del \cdot [\va \times (\del \times \vb)] = (\del \times \va) \cdot (\del \times \vb) - \va \cdot \del \times (\del \times \vb).
\end{gather}
Further use that $\del \times (\del \times \vb) = \del (\del \cdot \vb) - \del^2 \vb$ [\cite{book:korn}, Sec.~5.5-2]; thus \Eq{eq:ap1} and a symmetric expression for the third term on the right-hand side of \Eq{eq:ap2} can be put in the form
\begin{gather}
\del \cdot [\va \times (\del \times \vb)] = (\del \times \va) \cdot (\del \times \vb) - (\va \cdot \del) (\del \cdot \vb) + \va \cdot \del^2 \vb, \\
\del \cdot [\vb \times (\del \times \va)] = (\del \times \vb) \cdot (\del \times \va) - (\vb \cdot \del) (\del \cdot \va) + \vb \cdot \del^2 \va.
\end{gather}
Substitution of these into \Eq{eq:ap2} and then \Eq{eq:ap2} into \Eq{eq:ap3} yields
\begin{gather}\label{eq:ap4}
\{\va, \vb\} = \del^2 (\va \cdot \vb) - \va \cdot \del^2 \vb - \vb \cdot \del^2 \va - 2 (\del \times \va) \cdot (\del \times \vb).
\end{gather}
In Cartesian coordinates \Eq{eq:ap4} finally rewrites as [\cite{book:korn}, Sec.~5.5-5]
\begin{gather}\label{eq:cmtrab}
\{\va, \vb\} = 2 (\del A_j) \cdot (\del B_j) - 2 (\del \times \va) \cdot (\del \times \vb),
\end{gather}
where summation over repeated indexes is assumed.

\section{Pressure terms in the density equation}
\label{app:pressure}

\subsection{General case}

In this appendix we find $\del\cdot\vPi$ as a function of $n$ to the first order in $\epsilon$. Start off from \Eq{eq:pidef} to get
\begin{gather}\label{eq:divPi0}
\del\cdot\vPi \approx \frac{1}{\sme\N} \left(
i n k_j\,\frac{\pd P_{j\ell}}{\pd x_\ell} 
+ i \sngrad_j k_\ell \tpc_{j\ell} 
- \frac{\pd^2 \tpc_{j\ell}}{\pd x_j \pd x_\ell}
\right).
\end{gather}
The first term here is already of the sought form, and, since $\vngrad$ is of order $\epsilon$, the second term is expressed using
\begin{gather}\label{eq:aux3}
\tpc_{j\ell} \approx n \left(
\tPc_{j\ell} +
\tPc_{j s}\,\frac{k_sk_\ell}{k^2} +
\tPc_{\ell s}\,\frac{k_jk_s}{k^2}
\right),
\end{gather}
as obtained from the homogeneous stationary plasma approximation [\Eqsc{eq:uodep}{eq:parv}]. However, for calculating the third term \Eq{eq:aux3} is not sufficiently accurate \cite{foot:error}, so $\pd^2\tpc_{j\ell}/\pd x_j \pd x_\ell$ is found as follows.

First, take $\pd^2/\pd x_j \pd x_\ell$ of \Eq{eq:odep}, neglecting the terms of order higher than $\epsilon$. This yields
\begin{multline}\label{eq:aux2}
\frac{\db}{\dbdt}\left(
\frac{\pd^2 \tpc_{j \ell}}{\pd x_j \pd x_\ell}
\right) + 3 \tPc_{j \ell}\,\frac{\pd (\del \cdot \vu)}{\pd x_j \pd x_\ell} = \\
k_\ell n \Omega \left(
3\,\frac{\pd\tPc_{j \ell}}{\pd x_j} + \frac{k_\ell k_s}{k^2}\, \frac{\pd \tPc_{js}}{\pd x_j}   + \frac{3k_j k_s}{k^2}\,\frac{\pd\tPc_{j s}}{\pd x_\ell}
\right) 
+ k_j k_\ell \left( \tpc_{j \ell} \tWc_{ss} + 2 \tpc_{j s} \tWc_{\ell s} + 2 \tWc_{j s} \tpc_{s \ell} \right).
\end{multline}
The second term on the left-hand side allows an alternative representation via
\begin{gather}\label{eq:aux1}
\frac{\pd}{\pd x_j \pd x_\ell}\left(\del \cdot \vu\right) =  
n \Omega \sngrad_sk_s\,\frac{k_j k_\ell}{k^2} - \frac{\pd^2}{\pd x_j \pd x_\ell}\left(\frac{\db n}{\dbdt}\right)
\end{gather}
[see \Eq{eq:oden}], and the latter term in \Eq{eq:aux1} also equals
\begin{gather}
\frac{\pd^2}{\pd x_j \pd x_\ell}\left(\frac{\db n}{\dbdt}\right) = 
\frac{\db}{\dbdt}\left(\frac{\pd^2 n}{\pd x_j \pd x_\ell}\right)
- n \left(k_j k_s \tWc_{s \ell} + k_\ell k_s \tWc_{sj}\right).
\end{gather}
Hence \Eq{eq:aux2} can be put in the form
\begin{gather}\label{eq:Psi0}
\frac{\db}{\dbdt}\left(
\frac{\pd^2 \tpc_{j \ell}}{\pd x_j \pd x_\ell} - 3 \tPc_{j \ell}\,\frac{\pd^2 n}{\pd x_j \pd x_\ell}
\right) =  \Psi, \\
\Psi =  k_\ell n \Omega \left(
3\, \frac{\pd \tPc_{j \ell}}{\pd x_j} + \frac{k_\ell k_s}{k^2}\, \frac{\pd\tPc_{j s}}{\pd x_j} + 3\sN\,\frac{k_j k_s}{k^2}\, \frac{\pd \tTc_{j s}}{\pd x_\ell}
\right) + \delta \Psi,
\end{gather}
and $\delta \Psi$ is given by
\begin{gather}
\delta \Psi = 3 n k_j k_\ell\, (\dbt \tPc_{j \ell}) - 3 \tPc_{j \ell} n (k_j k_s \tWc_{s\ell}+k_\ell k_s \tWc_{sj}) + k_j k_\ell 
(\tpc_{j \ell}\tWc_{ss}+2\tpc_{js}\tWc_{\ell s}+2\tWc_{js}\tpc_{s\ell}).
\label{eq:aux4}
\end{gather}

Using the slow component of \Eq{eq:Pe} in the form
\begin{gather}\label{eq:slowp}
\dbt \tPc_{j \ell} = - \tPc_{j \ell} \tWc_{ss} - \tPc_{j s} \tWc_{\ell s} - \tPc_{s\ell} \tWc_{j s}
\end{gather}
and also \Eq{eq:aux3}, rewrite $\delta \Psi$ as
\begin{gather}
\delta \Psi = - 4n \vk \cdot (\tW \tG\tP \vk), \quad
\tG = \tI - \vk\vk/k^2.
\end{gather}
Since $\Psi$ is a rapidly oscillating function with a slow envelope of order $\epsilon$, \Eq{eq:Psi0} is integrated as
\begin{gather}\label{eq:Psi1}
\frac{\pd^2 \tpc_{j \ell}}{\pd x_j \pd x_\ell}  = 3 \tPc_{j \ell} \,
\frac{\pd^2 n}{\pd x_j \pd x_\ell}  + \frac{i \Psi}{\Omega}.
\end{gather}
Substitute \Eq{eq:Psi1} for the third term in \Eq{eq:divPi0}, together with \Eq{eq:aux3} for the second term; then one gets
\begin{gather}\label{eq:divpi}
\del \cdot \vPi = - \tCc_{j \ell}\,\frac{\pd^2 n}{\pd x_j \, \pd x_\ell}
-ik_\ell n \,\frac{\pd \tCc_{s\ell}}{\pd x_j} 
\left(\delta_{js} + \frac{k_jk_s}{k^2}\right) 
+ \frac{in}{3}\left(\frac{4}{\Omega}\,\vk\tW-\vngrad\right)\cdot \tG\tC \vk.
\end{gather}
Since the wave is assumed propagating along a local principal axis of the temperature tensor \cite{foot:notlang}, one has $\tG\tC \vk = \tG\vk \times \const$. Yet $\tG \vk \equiv 0$, so \Eq{eq:divpi} is simplified, and, using $ink_\ell \approx \pd n/\pd x_\ell$, one finally obtains
\begin{gather}\label{eq:divPi}
\del \cdot \vPi = - \tCc_{j \ell}\,\frac{\pd^2 n}{\pd x_j \, \pd x_\ell}
-\left(\delta_{js} + \frac{k_jk_s}{k^2}\right)\frac{\pd \tCc_{s\ell}}{\pd x_j}\,\frac{\pd n}{\pd x_\ell}.
\end{gather}
\end{widetext}

\subsection{Isotropic temperature}

For the isotropic temperature case, \Eq{eq:divPi} gives \cite{foot:parker}
\begin{gather}\label{eq:divPiiso}
\del \cdot \vPi = - 3\svte^2 \del^2 n - 6 \del n \cdot \del \svte^2,
\end{gather}
which as well can be obtained by substituting Eq.~(80) in Sec.~IV.3 of \Ref{ref:bernstein60} into our \Eq{eq:divPi0}. 

Alternatively, \Eq{eq:divPiiso} can be derived using a phenomenological adiabatic law (cf., \eg \Ref{ref:motz66}, \RefSec{book:allis}{Sec.~5.1}, \RefSec{book:stix}{Sec.~3.5})
\begin{gather}\label{eq:adiab}
(\pd_t + \vVe \cdot \del) (p_e N_e^{-\gamma}) = 0,
\end{gather}
with $p_e$ being the electron scalar pressure, including the slow and the quiver parts: $p_e = p + \soscp$. Here ${\gamma = 3}$ (corresponding to one-dimensional adiabatic oscillations~\cite{book:landau5}) is an extrapolation from the homogeneous plasma case, for which the exact solution is known from a more rigorous hydrodynamic treatment, like in our \Sec{sec:homogeneous} or \Refs{ref:tokatly99, tex:oberman60, ref:bernstein60}, or the complete kinetic treatment \citeSec{book:stix}{Chap.~8}, \cite{book:landau10}. To show this, introduce the plasma element Lagrangian displacement $\vxi$ such that \cite{ref:hayes70, ref:newcomb62, ref:peng75}
\begin{gather}
\soscN + \del \cdot (\vxi \N) = 0, \label{eq:xi1}\\
\soscp + \vxi \cdot \del p = \sme a^2 (\soscN + \vxi \cdot \del \N),
\end{gather}
where $a^2 = 3\svte^2$. Then, from \Eq{eq:pidef} with $\tP=p\tI$ and $\tp=\soscp\tI$, one gets
\begin{gather}
\vPi = - \del (n a^2) - [n \vQ + \del (\vxi \cdot \vQ)]/\N.
\end{gather}
Here $\vQ \equiv a^2 \del \N - \del p/\sme$ is of order $\epsilon$, and therefore one can take $\dbt \vxi \approx \vu$ (cf., \eg Eq.~(4.6) in \Ref{ref:newcomb62}), so $\vxi \approx i n \vk/k^2$, as flows from \Eq{eq:xi1}. Hence
\begin{gather}
\vPi \approx - \del (n a^2) - n\N^{-1} \tG \vQ,
\end{gather}
yielding
\begin{gather}
\del \cdot \vPi \approx - \del^2 (n a^2) - i n\N^{-1} \vk \tG \cdot \vQ.
\end{gather}
Using that $\vk \tG \equiv 0$, one finally obtains
\begin{gather}
\del \cdot \vPi \approx - a^2 \del^2 n - 2\del n \cdot \del a^2,
\end{gather}
which is equivalent to \Eq{eq:divPiiso}.

\section{Ponderomotive pressure}
\label{app:eos}

The conservation of the Langmuir wave action allows to calculate the effective stress tensor due to the wave, which is done as follows (see also \Refs{ref:dewar77, ref:kentwell87, ref:pitaevsky60}). For simplicity suppose homogeneous cold stationary plasma volume $\mc{V}$ and assume that it is adiabatically deformed as defined by an infinitesimal displacement field $\vxi(\vrr)$, resulting in the strain tensor
\begin{gather}
\tw = [(\del \vxi) + (\del \vxi)^\transpose]/2.
\end{gather}
Hence the stress tensor $\tsigma$ is found \cite{book:landau7}:
\begin{gather}\label{eq:sigmadef}
\tsc_{j \ell} = \frac{1}{\mc{V}}\,\frac{\pd \mcc{E}}{\pd \swc_{j \ell}},
\end{gather}
where $\mcc{E} =\mc{V} J \omega$ is the wave total energy inside $\mc{V}$. Because $\mc{V} J$ is conserved, \Eq{eq:sigmadef} rewrites as
\begin{gather}\label{eq:sig1}
\tsc_{j \ell} = \frac{J \omega_p}{2\N}\,\frac{\pd \N}{\pd \swc_{j \ell}},
\end{gather}
where we used that $\omega = \omega_p(\N)$. The density perturbation due to the strain is $\delta \N = - \N \del \cdot  \vxi$ [cf. \Eq{eq:xi1}]. Since $\del \cdot \vxi = \swc_{ss}$, and $\pd \swc_{ss}/\pd \swc_{j \ell} = \delta_{j \ell}$, this yields
\begin{gather}
\tsigma = - (J \omega_p/2)\,\tI.
\end{gather}
Therefore the stress due to the wave field is isotropic and appears as an effective pressure $p_E = J \omega_p/2$. (Thermal correction would also yield an anisotropic component to the wave stress tensor \cite{ref:kentwell87, ref:dewar77}.) Using \Eq{eq:coldj}, one then recovers the expression [\Eq{eq:peff}] for the ponderomotive pressure in cold plasma carrying a field $\voscE$ which oscillates at the frequency $\omega = \omega_p$ \cite{ref:kentwell87}.


\begin{thebibliography}{10}
\bibitem{foot:nonham}
The effects of non-Lagrangian phenomena like ionization or recombination are
  described in \Refs{ref:stepanov76, ref:biragov85, ref:stepanov96,
  ref:gildenburg93, ref:bakunov00}.
\bibitem{foot:anisot}
From \Eqsc{eq:slowp}{eq:slown} one has for the temperature tensor $[\partial_t
  + (\vv \cdot \del)] \tT = - (\tW\tT)-(\tW\tT)^\transpose$, where $\vv$ is the
  average velocity, and $\tW = \del \vv$.
\bibitem{foot:nonrel}
The wave is not purely longitudinal in inhomogeneous plasma at nonzero $\vv$
  \cite{ref:pikulin79}. Yet the effect of the wave magnetic field vanishes in
  the nonrelativistic limit, which is discussed here; hence the potential
  approximation.
\bibitem{foot:parker}
\Eqsc{eq:nmainiso}{eq:divPiiso} complete Eq.~(7) of \Ref{ref:mendonca09b} (yet
  see \cite{foot:nonham}) and correct \Ref{ref:parker64} in how the pressure
  terms appear when plasma is inhomogeneous.
\bibitem{foot:notlang}
Otherwise a wave cannot be locally approximated with a plane Langmuir wave
  found in \Sec{sec:homogeneous}. For details on geometrical optics of
  eigenmodes in anisotropic media see \Ref{book:kravtsov}, Secs.~4.2, 4.3.
\bibitem{foot:b}
An alternative case, not considered here, is compression due to external
  magnetic field.
\bibitem{foot:ion}
When nonlinear effects are taken into account, the electron ponderomotive force
  also enters the ion dynamics equations; see, \eg \Ref{ref:robinson97}.
\bibitem{foot:error}
This point is missed in Appendix~C of \Ref{ref:bernstein77}, where the
  conductivity of inhomogeneous nonstationary anisotropic unmagnetized plasma
  is calculated. Hence the corresponding corrections due to the temperature do
  not comply with PCT.
\end{thebibliography}

\end{document}